\documentclass[aps,prl, twocolumn,preprintnumbers,10pt,showpacs,superscriptaddress]{revtex4-1}
\usepackage[english]{babel}

\usepackage{dcolumn}
\usepackage{amssymb}   
\usepackage{pifont}    
\usepackage{makecell}
\usepackage{amsmath}
\usepackage{amsthm,amssymb}
\usepackage{amstext}
\usepackage{orcidlink}
\usepackage{listings}
\usepackage{graphicx,subfigure,setspace,soul}
\usepackage[table]{xcolor}
\usepackage{booktabs}

\definecolor{dkgreen}{rgb}{0,0.6,0}
\definecolor{gray}{rgb}{0.5,0.5,0.5}
\definecolor{mauve}{rgb}{0.58,0,0.82}
\definecolor{success}{RGB}{0,150,0}
\definecolor{failure}{RGB}{200,0,0}
\definecolor{pending}{RGB}{220,120,0}

\newcommand{\cmark}{{\color{success}\checkmark}}
\newcommand{\xmark}{{\color{failure}\ding{55}}} 

\usepackage{hyperref}

\newcommand{\nd}{\mathrm{d}} 

\allowdisplaybreaks

\begin{document}

\title{\boldmath Uncovering Singularities in Feynman Integrals via Machine Learning }
\author{Yuanche Liu\,\orcidlink{0009-0008-4604-1306}}
\email{liuyuanche@mail.ustc.edu.cn}	
\affiliation{School of Physical Sciences, University of Science and Technology of China, Hefei, Anhui 230026, China}

\author{Yingxuan Xu\,\orcidlink{0000-0001-6135-8864}}
\email{yingxuan.xu@kit.edu}	
\affiliation{Institute for Theoretical Particle Physics, KIT, Wolfgang-Gaede-Straße 1, 76131, Karlsruhe, Germany}

\author{Yang Zhang\,\orcidlink{0000-0001-9151-8486}}
\email{yzhphy@ustc.edu.cn}	
\affiliation{Interdisciplinary Center for Theoretical Study, University of Science and Technology of China, Hefei, Anhui 230026, China}
\affiliation{Peng Huanwu Center for Fundamental Theory, Hefei, Anhui 230026, China}
\affiliation{Center for High Energy Physics, Peking University, Beijing 100871, People’s Republic of China}

\preprint{} 
\date{\today}
\preprint{USTC-ICTS/PCFT-25-40, P3H-25-073, TTP25-034}
\begin{abstract}
We introduce a machine-learning framework based on symbolic regression to extract the full symbol alphabet of multi-loop Feynman integrals. By targeting the analytic structure rather than reduction, the method is broadly applicable and interpretable across different families of integrals. It successfully reconstructs complete symbol alphabets in nontrivial examples, demonstrating both robustness and generality. Beyond accelerating computations case by case, it uncovers the analytic structure universally. This framework opens new avenues for multi-loop amplitude analysis and provides a versatile tool for exploring scattering amplitudes.

\end{abstract}
\maketitle
\section{Introduction}
\label{sec:intro}

High-precision scattering amplitudes are crucial for testing the Standard Model at colliders and modeling gravitational waves from compact binaries. Upcoming experiments such as the HL-LHC, CEPC, FCC-ee, and third-generation gravitational-wave detectors will achieve unprecedented precision, demanding theoretical predictions of comparable accuracy, particularly in the form of accurate multi-loop scattering amplitudes.

Around a decade ago, obtaining precise predictions for two-to-three particle collider processes beyond next-to-leading order was widely considered infeasible. This changed with advances in evaluating complicated two-loop Feynman integrals and interpreting them in terms of Chen's iterated integrals. Key steps include deriving and solving differential equations for master integrals and assembling full amplitudes, often with finite-field techniques. In this context, the concept of the symbol alphabet and associated function spaces has become central for multi-loop studies \cite{Goncharov:2009lql, Goncharov:2010jf}. These tools capture the algebraic structure of iterated integrals, first explored by Chen in the 1970s \cite{Chen:1977oja}, which naturally arise in canonical-form differential equations~\cite{Henn:2013pwa} and can be expressed as nested \textrm{d-log} integrals.

The symbol alphabet not only encodes the algebraic structure of iterated integrals but also underpins modern amplitude techniques, including bootstrap methods for master integrals and full scattering amplitudes~\cite{Dixon:2011pw, Dixon:2013eka, Drummond:2014ffa, Chicherin:2017dob, Caron-Huot:2019vjl}. Several tools facilitate symbol analysis: \texttt{HyperInt} implements a reduction algorithm yielding a superset of Landau singularities~\cite{Panzer:2014caa}, \texttt{PLD.jl} and \texttt{SOFIA} with \texttt{Effortless} implementation compute singularities~\cite{Fevola:2023kaw,Fevola:2023fzn,Matijasic:2024too,Caron-Huot:2024brh, Correia:2025yao}, and \texttt{Baikovletter} attempts to reconstruct symbol alphabets via Baikov representations~\cite{Jiang:2024eaj}.

In this letter, we propose a symbolic-regression–based approach to uncover the singularity structure of multi-loop Feynman integrals. The framework combines efficient numerical evaluations with interpretable machine-learning techniques, enabling us to determine the canonical differential equations and to generate the complete symbol alphabet. Unlike previous machine-learning applications that focus mainly on accelerating IBP reductions~\cite{vonHippel:2025okr, Zeng:2025xbh, Song:2025pwy}, our method directly targets the analytic structure, requiring neither prior knowledge of singularities nor computationally costly analytical reduction steps. We demonstrate the approach on nontrivial multi-loop examples, showing how all symbol letters—including square-root–type structures—can be systematically identified.

\section{Canonical Differential Equations and Symbol Alphabets}
\label{sec:can}
\subsection{From Feynman Integrals to Canonical Basis}
A generic $L$-loop Feynman integral can be written as
\begin{gather}
\int\prod_{j=1}^L \frac{d^D l_j}{i \pi^{D/2}}\frac{1}{\prod_{i=1}^n D_i^{\alpha_i}},
\label{Feynman integral}
\end{gather}
here $L$ is the number of loops, and $\alpha_i$ can be positive or negative integers, representing the denominator and numerator factors. For simplicity, a Feynman integral can be denoted as $I[\alpha_1, \dots, \alpha_n]$.

The formal IBP relation for Feynman integrals in momentum space is~\cite{Tkachov:1981wb,Chetyrkin:1981qh},
\begin{gather}
0=\int\prod_{j=1}^L \frac{d^D l_j}{i \pi^{D/2}}\sum_{k=1}^L\frac{\partial}{\partial l_{k}^\mu}\frac{v_k^\mu }{\prod_{i=1}^n D_i^{\alpha_i}},
\label{IBP-relation}
\end{gather}
where $v^{\mu}$ is an arbitrary vector built from the loop momenta or the external momenta. With the IBP identities in \eqref{IBP-relation}, a large linear system of Feynman integrals can be generated. By Gaussian elimination, the basis of a given integral family can be found. The basis is called master integrals. The existence of master integrals is proved in Ref.~\cite{Smirnov:2010hn}.

The value of a standard Feynman integral depends only on scalar products of external momenta and masses. Taking derivatives with respect to these invariants yields partial differential equations that can be used to evaluate the integrals~\cite{Remiddi:1997ny,Gehrmann:1999as}. It was shown in~\cite{Henn:2013pwa,Henn:2014qga} that, with a suitable choice of master integrals, the dimensional regulator $\epsilon$ can factor out of the system. The equations then admit a simple solution in an $\epsilon$-expansion\cite{e-collaboration:2025frv}. The corresponding choice of master integrals is called uniform transcendental (UT) basis. Such systems are known as canonical differential equations. They take the universal form
\begin{gather}
    d\vec{I}=\epsilon (d A) \vec{I} \,,
\label{DE}
\end{gather}
where the matrix $A$ encodes the analytic structure of the integrals. In the letter, we consider a broad class of Feynman integrals with polylogarithm expressions whose differential equation can be further decomposed as,
\begin{gather}
    A=\sum A_k \log S_k \,,
\label{dlog}
\end{gather}
with rational number matrices $A_k$ and algebraic kinematic functions $S_k$, known as symbol letters. This covers a wide range of cases that appear in calculations from the Standard Model and gravitational-wave physics, and the extension beyond algebraic letters is left for future investigation.

\subsection{Symbol Alphabet from Canonical Differential Equations}

The solution of Eq.~\eqref{DE} can be formally expressed as an iterated integral:  
\begin{equation}
f = \int \nd\log S_n \cdots \int \nd\log S_2 \int \nd\log S_1 \,,
\end{equation} 
where the integration is performed along a chosen path in kinematic space.  
Each $S_i$ is a function of external momenta or masses, i.e.\ the symbol letters appearing in~\eqref{dlog}.  
The symbol of such an iterated integral is given by~\cite{Goncharov:2009lql,Goncharov:2010jf}
\[
\mathcal{S}(f) = S_1 \otimes S_2 \otimes \cdots \otimes S_n \,,
\]  
from which the full function can be reconstructed, up to boundary terms.

Once the symbol letters in Eq.~\eqref{dlog} and the weight-zero boundary values are known, the symbol of the solution follows directly.  A basis for the relevant function space can then be built.  
An ansatz for the full result is then proposed, with its coefficients determined by imposing physical constraints.  
This strategy has proven powerful in practice, and has been successfully applied to scattering amplitudes~\cite{Dixon:2011pw,Dixon:2011nj},  
form factors~\cite{Brandhuber:2012vm},  
soft anomalous dimensions~\cite{Li:2016ctv,Almelid:2017qju}, Wilson loop~\cite{Carrolo:2025pue} as well as to many classes of Feynman integrals~\cite{Henn:2018cdp,He:2021eec}.

However, the analytic construction of canonical differential equations is computationally challenging. Even with full analytical expressions in CDE matrices, it is still hard to reconstruct symbol letters from it. As a result, this natural route --- analytic derivation of canonical differential equations and direct extraction of symbol letters is often impractical in practice. In the next section, we address this problem by employing modern symbolic regression frameworks, in particular \textsc{PySR}, to systematically search for and identify candidate symbol letters in demand of \textbf{only} numerical differential equations.

\section{Symbolic Regression Frameworks and PySR}
\label{sec:reg}
Symbolic regression aims to identify analytic expressions that capture relations between physical variables, with the functional form itself treated as an unknown. Mathematically, 
considering a grammar $\mathcal{G}$ of admissible operators (e.g.,
$+,-,\times,\sin,\exp$), variables, and constants, which generates a discrete space 
of symbolic structures $s \in \mathcal{S}$. Each structure admits continuous 
parameters $\theta \in \mathbb{R}^{k(s)}$, yielding functions 
$f_{s,\theta}:\mathbb{R}^d\to\mathbb{R}$. The task is thus a mixed discrete–continuous optimization: fitting continuous parameters $\theta$ within each structure while selecting the structure $s$ that achieves an optimal balance between accuracy and complexity. Formally, this can be written as 
\begin{equation}
  (s^\star,\theta^\star) = \arg\min_{s\in\mathcal{S}} 
\left\{ \min_{\theta\in\mathbb{R}^{k(s)}} 
\,\mathcal{L}_D(f_{s,\theta}) + \lambda\,\mathcal{C}(s) \right\},
\end{equation}
where $\mathcal{L}_D$ is a data loss (e.g., the squared predictive error with respect to physical measurements) and $\mathcal{C}(s,\theta)$ is a complexity penalty that encodes a preference for simple expressions. From a physics perspective, this 
separation reflects common practice: once a law is hypothesized in form, its 
constants are determined numerically. Moreover, symbolic regression naturally admits a multi-objective interpretation, in which predictive error $E(s,\theta)$ and model complexity $C(s)$ are treated as distinct optimization objectives. The Pareto front 
of non-dominated solutions provides a spectrum of candidate laws. It spans the simplicity–accuracy trade-off, from simple yet approximate to complex but precise, offering a principled balance between empirical adequacy and interpretability.

In practice, most SR frameworks rely on evolutionary search, where populations of candidate formulas are iteratively modified and selected for fitness~\cite{schmidt2009distilling, Song:2025odk}. Alternative approaches include reinforcement-learning strategies and neural methods (e.g.\ Deep Symbolic Regression and transformer-based decoders), which guide the search via learned policies or pre-trained large models~\cite{biggio2021deep}.

\subsection{\textsc{PySR}: A High-Performance Symbolic Regression Engine}

\textsc{PySR}~\cite{cranmer2023PySR} is an open-source Python library 
(with a Julia backend) designed for practical symbolic regression. 
The high-performance \texttt{SymbolicRegression.jl} engine, used by \textsc{PySR}, implements a multi-population evolutionary algorithm with periodic migrations to promote population diversity. Crucially, 
\textsc{PySR} augments genetic programming with a specialized constant-optimization step: numeric constants in each generated symbolic expression are optimized using gradient-free methods from \texttt{Optim.jl}. This ``Evolve–Simplify–Optimize'' loop makes \textsc{PySR} much more reliable than naive Genetic Programming, which relies solely on random constant mutations.

\begin{widetext}

  \begin{figure}[htbp]
    \centering
    \includegraphics[width=.75\textwidth]{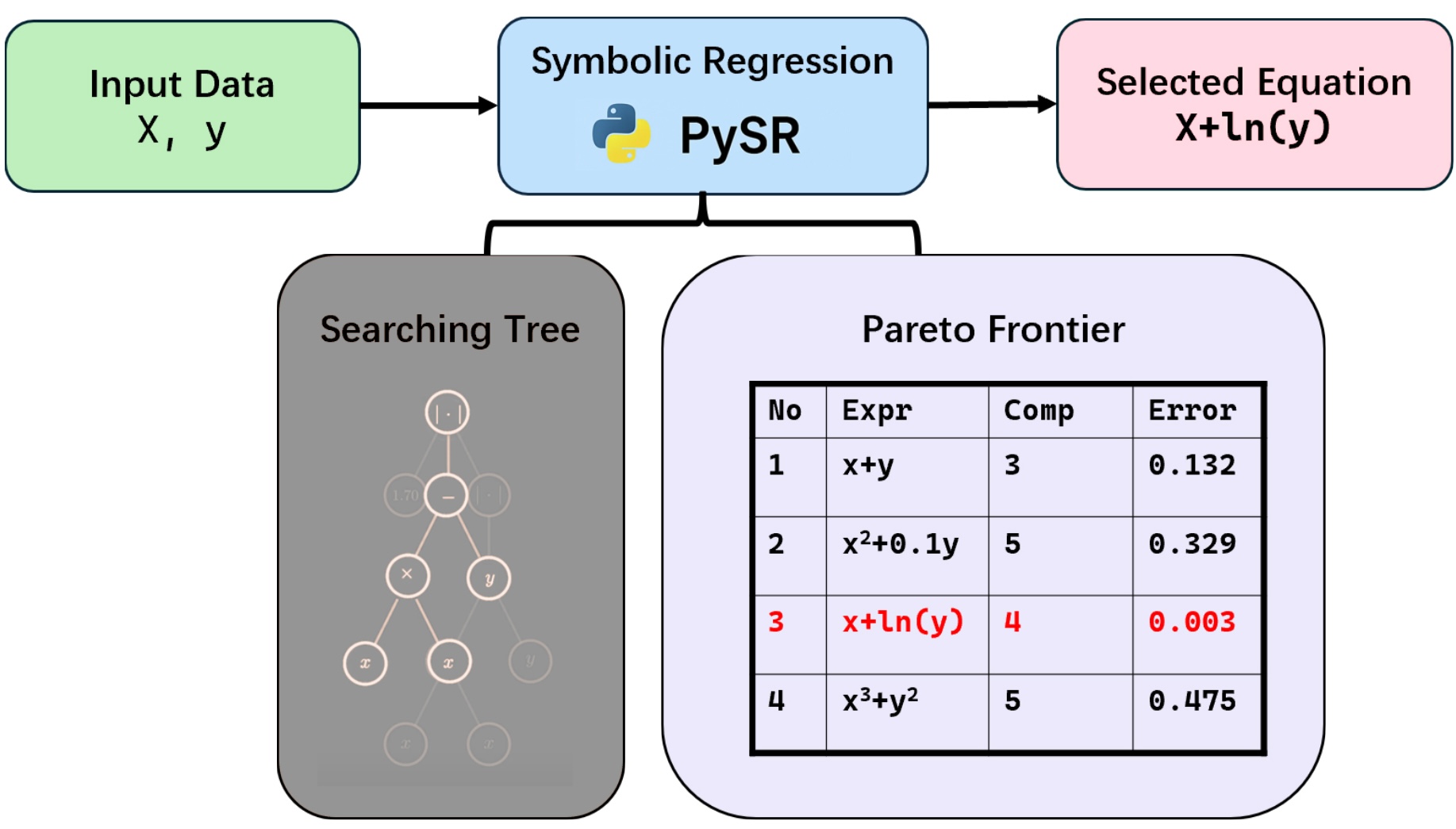}
    \caption{\textsc{PySR} Workflow. Dataset characters will be learned by evolution of searching tree, which combines operators and operands to a test expression. After considering the Pareto Frontiers of expressions in respect to their complexity, size, and accuracy, \textsc{PySR} will return several best results to its user.}
    \end{figure}
\end{widetext}

\textsc{PySR} combines distributed evolutionary search with efficient 
constant optimization, producing accurate and interpretable symbolic models. Its design makes it well-suited for methodological innovation and practical use in physics research, including applications to ODE/PDE discovery. Compared with other symbolic regression tools, \textsc{PySR} offers a favorable balance: Frameworks like \texttt{gplearn} are slower and less precise due to lack of constant optimization, while 
\textsc{C++} libraries like Operon\cite{10.1145/3377929.3398099} are extremely fast but less accessible and hard to extend. Neural-network-based approaches (e.g.\ AI-Feynman\cite{udrescu2020ai,udrescu2020ai2}) can be powerful but require pre-training or large datasets. By contrast, \textsc{PySR} is both efficient and accessible, making it especially useful for physics applications where closed-form models are essential.


\subsection{Build Symbol Letters From Numerical Evaluations}

Most symbolic regression methods aim to seek higher precision and interpretable symbolic expressions to describe meaningful relationships between physical observables and to predict results for unknown points in parameter space. These symbolic relations are not necessary to be absolutely correct. In our work, however, we generates symbolic results through a "overfitting-directed" tuning procedure, to obtain completely accurate symbolic expressions.

The workflow begins with the construction of a canonical basis for a given family of Feynman integrals. Numerical IBP reductions are performed with \textsc{Kira}~\cite{Maierhofer:2017gsa,Klappert:2020nbg,Lange:2025fba} to obtain CDE matrices at multiple kinematic points. (Since \textsc{PySR} performs regression with floating-point numbers, we truncate the rational coefficients from IBP reductions to 30 significant digits, balancing efficiency and accuracy.) Given up to $\mathcal{O}(100)$ pairs $(x_i,{A}(x_i))$, \textsc{PySR} searches for analytic candidates of the matrix elements and iteratively optimizes them. The resulting expressions are verified to exhibit a \textrm{d-log} form. Exponentiating and factorizing these entries yields a set of symbol letters, which constitute the candidate complete symbol alphabet for the integral family. The following figure \ref{fig:workflow} visualizes our framework.

To illustrate the workflow, consider a single-variable system. In this case the CDE matrix elements take the total-derivative form of Eq.~\eqref{dlog}. Numerical IBP reductions at points $x_1,\dots,x_n$ provide matrices $\bm{A}(x_1),\dots,\bm{A}(x_n)$. The task then reduces to an indefinite integration problem: given $\bm{A}(x)$ at discrete points, determine $f(x)$ such that $\mathrm{d}f/\mathrm{d}x=\bm{A}(x)$. Once a closed form of $f(x)$ is found, the symbol letters can be read off, yielding the complete alphabet.

\begin{widetext}

  \begin{figure}[htbp]
    \label{fig:workflow}
    \centering
    \includegraphics[width=.95\textwidth]{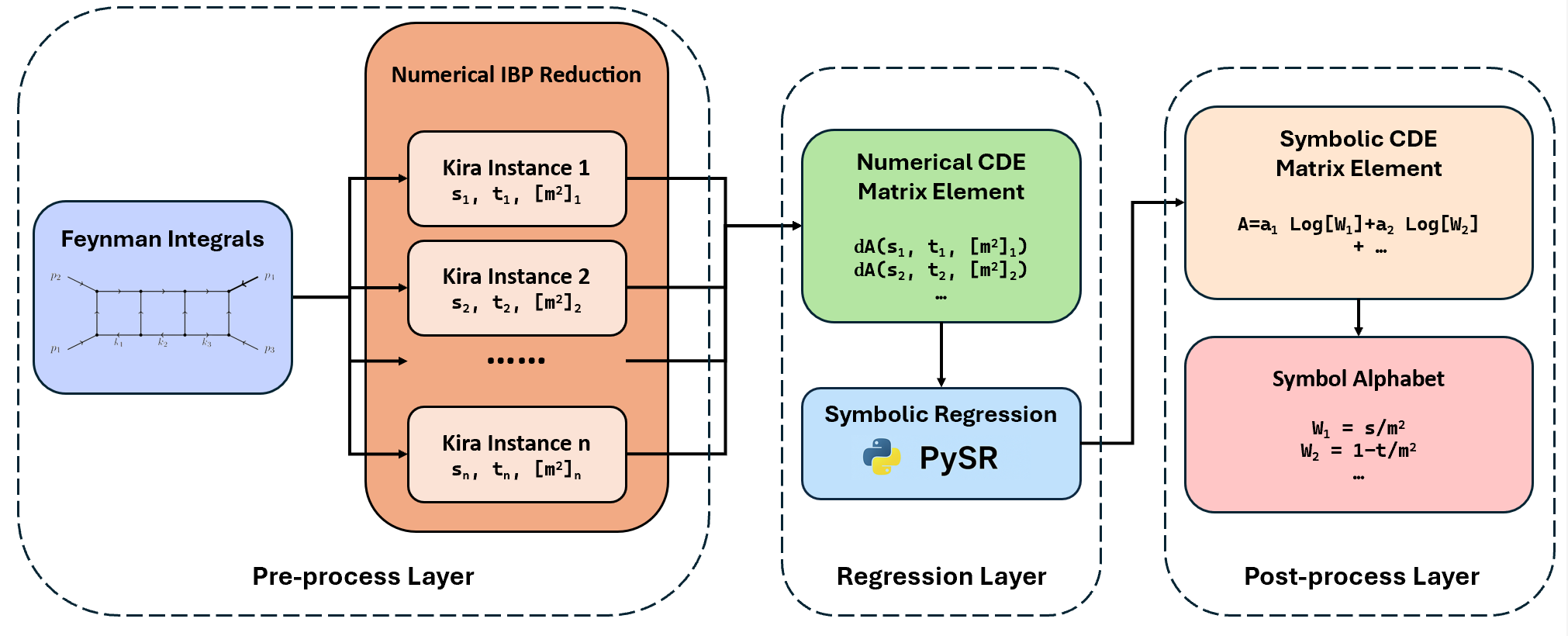}
    \caption{The workflow consists of three layers. In the Pre-processing Layer, Feynman integrals of a given family are analyzed and IBP reductions are performed at multiple numerical points. These results serve as input for the Regression Layer, where symbolic regression constructs the CDE matrix. In the Post-processing Layer, the resulting symbolic expressions are exponentiated and factorized, and all candidate symbol letters are collected to assemble the complete symbol alphabet.}
    \end{figure}
\end{widetext}

An addition knowledge on CDEs restrict the functions that appear in $f$. To be precise, there are only \texttt{log} and (sometimes) \texttt{sqrt} functions, and the \texttt{sqrt} functions survive only inside \texttt{log} functions. \textsc{PySR} also support a user-defined constraints on (nested) function structures. For example, the restrictions mentioned above can be achieved by defining an arbitrary \texttt{PySRRegressor} (see Listing \ref{code:2} for a detailed code template).

During the initial epochs, \textsc{PySR} may generate many candidate expressions with non-negligible errors (By default, \textsc{PySR} use Root Mean Square Error(RMSE) to estimate the regression error. initial errors may reach $10^{-1} \sim 10^{-3}$). One should carefully tune Hyper-parameters to avoid underfitting. After approximatly 40 epoches in less than one hour (depends on the complexity of target expressions),  \textsc{PySR} identifies expressions formally equivalent to the target, with errors reduced to $\sim 10^{-30}$, consistent with the precision of the input. This means \textsc{PySR} has find a proper expression equivalent to the correct answer, and corresponding symbolic expressions can be exponentiated and factorized to extract the letters appearing in each entry. One can now stop the searching to avoid overfitting and extract symbol letters from the results, and this can be automated by tuning Hyper-parameter \texttt{early\_stop\_condition}.

\section{Examples}
\label{sec:exm}

In this section we present some cases to display our workflow. We also provide a sheet recording our trials on Feynman Integrals families with different number of scales and  loops.

\subsection{Planar three-loop four-point one-mass integrals}
\label{3l4p-example}
Following the notation and definitions in Ref.~\cite{Henn:2023vbd}, we focus on the three-loop four-point Feynman diagram as follows,
\begin{figure}[ht]
    \centering
    \includegraphics[width=0.4\textwidth]{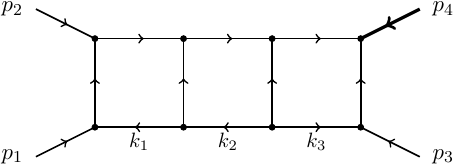} 
    \caption{Tripple box Feynman diagrams.}
    \label{fig:diagram}
\end{figure}

Using $s=(p_1+p_2)^2$, $t=(p_1+p_3)^2$, $m^2=p_4^2$, we introduce the dimensionless variables
\begin{equation}
  x = \dfrac{s}{m^2}, \quad y = \dfrac{t}{m^2} 
\end{equation}
thus the symbol letters depend only on $x$ and $y$.

This integral family contains 83 master integrals. The numerical reduction at a sample point, e.g. $s = 7, t=-9/11, p_4^2 = 3$, containing target integrals with denominators up to degree 11, numerators up to degree 3, and with a maximum of 4 double propagators, can be completed in less than 30 minutes on an Intel(R) Core(R) i9-13950HX CPU @ 5.30GHz with 12 of its 32 cores.

Based on the UT basis shown in Ref.\cite{Henn:2023vbd}, the numerical canonical differential equations with regarded to the kinematic variables can be constructed. We perform numerical IBP reductions in 200 different kinematic points and build corresponding numerical CDE matrices. For example, one entry of the analytical CDE matrix is
\begin{equation}
  \label{eqs:Res1}
  f(x,y) = \dfrac{14}{15} \log(1 - x) - \dfrac{2}{5} \log\dfrac{1 - x - y}{1 - x} + \dfrac{2}{5} \log(y)
\end{equation}

Proceed this certain entry of numerical CDE matrices with our fine-tuned PySR regressor, one may obtain the following result after about 50 epoches:

\begin{table}[htbp]
  \centering
  \setlength{\tabcolsep}{6mm}
  \renewcommand{\arraystretch}{1.2}

  \begin{tabular}{c|cc}
    Complexity & Loss     & Equation  \\\hline
    \dots      & \dots & \dots \\
    $47$         & $0.892719$ & $f = f_1$    \\
    \rowcolor[gray]{0.93}
    $52$         & $5.70\mathrm E-29$ & $f = f_2$    \\
    \dots      & \dots & \dots 
    \end{tabular}
  \caption{Part of Symbolic Regression results obtained from \textsc{PySR} output.}
  \label{tab:multiloop}
\end{table}
\noindent where $f_1$ and $f_2$ are complicated expressions. The highlighted row is obtained from  $f_2$ and turns out to be as precise as input numerical data. One can treat this as target result, and will get a simplified symbolic expression for $f_2$:

\begin{equation}
  \label{eqs:simpRes1}
  f_2 = \dfrac{4}{3} \log(1-x) - \dfrac{2}{5}\log (1-x-y) + \dfrac{2}{5} \log(y)
\end{equation}
One can verify the equivalence between \eqref{eqs:Res1} and \eqref{eqs:simpRes1}. From \eqref{eqs:simpRes1} one find three distinguished letters in this entry: 
\begin{equation}
  \label{eqs:3l4pRes1}
  W_1 = 1-x, \quad W_2 = 1-x-y, \quad W_3 = y
\end{equation}

Applying these processes to all CDE matrix elements, complete symbol alphabet is as follows: 
\begin{equation}
  \label{eqs:3l4pRes2}
  W_i = \{x, 1-x, y, 1-y, x+y, 1-x-y\}
\end{equation}

which is in good agreement with that shown in Ref.\cite{Henn:2023vbd}.

\subsection{Non-planar two-loop three-point Feynman integrals}
In this section, we consider a non-planar two-loop three-point Feynman diagram which has been evaluated in Ref.~\cite{vonManteuffel:2017hms}.

\begin{figure}[ht]
    \centering
    \includegraphics[width=0.2\textwidth]{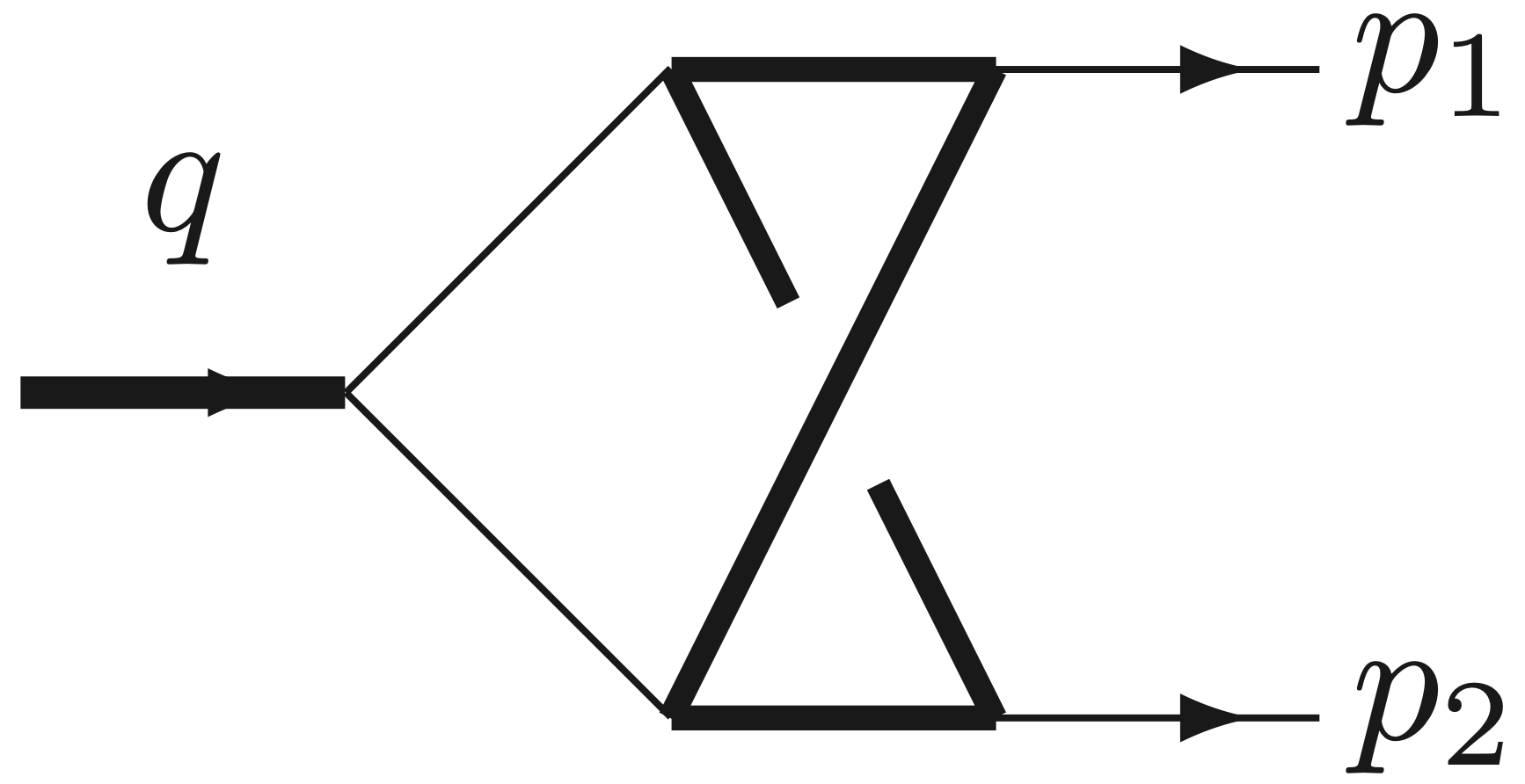} 
    \caption{Non-planar two-loop three-point Feynman diagram.}
    \label{fig:diagram}
\end{figure}

The integral basis can be expressed in terms of complete elliptic integrals and polylogarithms. Consequently, unlike the case in Sec.~\ref{3l4p-example}, which involves only rational functions, several square-root letters appear in the symbol alphabet of this family.

Based on the diagram, four of the six propagators are massive (thick lines), while only one of the three external legs is oﬀ-shell, i.e., $p_1^2=p_2^2=0$ and $q^2=(p_1+p_2)^2=s$. For convenience, one can introduce the mass ratio, 
\begin{gather}
x = -\frac{s}{m^2}.
\end{gather}

Using the same workflow as in Sec.~\ref{3l4p-example}, we identified the following five symbol letters:
\begin{gather}
    l_1 = \sqrt{x}, \quad l_2 = \tfrac{1}{2}\!\left(\sqrt{x} + \sqrt{x+4}\right), \\\nonumber
    l_3 = \sqrt{x+4}, \quad l_4 = \tfrac{1}{2}\!\left(\sqrt{x} + \sqrt{x-4}\right), \quad l_5 = \sqrt{x-4}.
\end{gather}
These results are in agreement with Ref.~\cite{vonManteuffel:2017hms}.

Table \ref{tab:result} lists all tested cases. Our workflow successfully reconstructs the complete symbol alphabet for most families of Feynman integrals with up to four external momenta. For integrals with more scales, it recovers almost all even letters, while the remaining ones with complicated square-root structures can be derived in the canonical basis, and odd letters can then be built. Then we successfully obtain the complete alphabet from the cutting-edge planar 3-loop 5-point Feynman integral family~\cite{Liu:2024ont}.

\begin{widetext}

  \begin{table}[htbp]    
    \label{tab:result}
    \centering
    \setlength{\tabcolsep}{6mm}
    \renewcommand{\arraystretch}{2}
    \rowcolors{2}{gray!10}{white}
  
    \begin{tabular}{>{\bfseries}c|ccccc}
      \#Loop $\backslash$ Family  & \makecell{{1 Scale} \\ 
\texttt{3p1m,4p0m}} & \makecell{2 Scales \\ \texttt{4p1m}} & \makecell{3 Scales\\ \texttt{4p2m}} & \makecell{5 Scales\\ \texttt{5p0m}} & \makecell{5+ Scales\\ \texttt{5p1m, 6p0m}} \\
      \hline
      1 & \cmark\cmark & \cmark\cmark & \cmark\cmark & \cmark& \cmark \\
      2 & \cmark\cmark & \cmark\cmark & \cmark\cmark & \cmark & \xmark \\
      3 & \cmark\cmark & \cmark\cmark & \cmark\cmark & \cmark & —— \\
      4 & \cmark\cmark & —— & —— & —— & —— \\
    \end{tabular}
  
    \caption{Trials on multi-loop, multi-scale integral families.
\cmark\cmark: complete symbol alphabet reconstructed;
\cmark: all even letters obtained, remaining ones (odd letters) can be constructed manually;
\xmark: some letters not found;
——: not tested because of incomplete reference.
An $x$-p $y$-m integral has $x$ external momenta, of which $y$ are massive. }
    \label{tab:multiloop}
  \end{table}
  
\end{widetext}

\section{Conclusion}

In this letter, we introduced a machine-learning framework based on PySR to extract the complete symbol alphabet of multi-loop Feynman integrals. By directly targeting the analytic structure rather than relying on explicit integral representations or singularity analysis, our approach circumvents the limitations of existing methods, which often struggle with odd letters or collapse in the presence of square-root and elliptic structures. Its effectiveness was demonstrated through several nontrivial examples, underscoring both its robustness and its broad applicability across different families of integrals.

Looking forward, this framework opens multiple avenues for exploration. It provides a systematic and transparent route to analytic structures without prior singularity knowledge or costly IBP reductions and reconstructions, and it suggests a pathway toward the automated discovery of symbol alphabets in higher-loop and multi-leg amplitudes. Beyond symbol extraction, the same strategy could be extended to bootstrap program or integrability to accelerate the analytic determination of hidden algebraic structures. More broadly, our work highlights how machine learning can serve not only as a computational accelerator but also as a conceptual guide in uncovering the analytic structures underlying quantum field theory.

\section{Acknowledgment}
We thank Miles Cranmer, the author of PySR, for his assistance on the usage of PySR. We also Thank Tianji Cai for discussion on symbolic regressions and application of machine learning on physics. Yuanche Liu is  supported by National Natural Science Foundation of China (NSFC) through Grant No. 124B1014. The research of Yingxuan Xu is supported by the Deutsche Forschungsgemeinschaft (DFG, German Research Foundation) under grant 396021762 - TRR 257. Yang Zhang is supported by NSFC through Grant No. 12575078 and 12247103.

\appendix

\bibliographystyle{h-physrev}
\bibliography{biblio.bib}

\section{Supplemental Material}\label{sec:supp}

In this appendix we list all supplemental materials attached to this letter, and explain their usage in our workflow.

\begin{itemize}
    \item \texttt{regression.py}:
    Main program for calling PySR to perform symbolic regressions. This program corresponds to the three-loop four-point one-mass example shown in Sec.~\ref{3l4p-example}.  
    For other cases---especially when the symbol letters involve square roots or more complicated algebraic structures---the options of \texttt{PySRRegressor} and several hyperparameters may need manual adjustment.  
    Further details can be found in the PySR documentation.  
    \item \texttt{cord\_1.txt, cord\_2.txt}:
    Numerical results for specific CDE entries in the 3l4p1m example.
 Containing 200 random kinematic points $(x, y)$ with 30-digit precision, which can be used as input for numerical CDE evaluations. 
    \item \texttt{dinv\_1.txt, dinv\_2.txt}:
    Data files consistency with the 3l4p1m example. Including 200 values truncated to 30-digit precision, obtained from numerical IBP reductions.
    \item \texttt{hall\_of\_fame.csv}:
    The search was continued until an evident loss gap from $\sim10^{-1}$ to $\sim10^{-30}$ was achieved. One can also find the full expressions of $f_1$ and $f_2$ to verify our results displayed in \ref{3l4p-example}.
\end{itemize}

\section{Options and Tuning of PySR Regressor}\label{sec:code}

For single variable cases, PySR natively supports differential operators, enabling indefinite integration. One may set a ``Template '' of target function:

\begin{lstlisting}[language=Python, caption={An arbitrary function template for indefinite integration in PySR},captionpos=b]
  from pysr import PySRRegressor, TemplateExpressionSpec

  expression_spec = TemplateExpressionSpec(
      expressions=["f"],
      variable_names=["x"],
      combine="df = D(f, 1); df(x)",
  )
\end{lstlisting}

PySR interprets the input as numerical samples of \texttt{df(x)}, the derivative of $f(x)$, and searches for $f(x)$ via symbolic regression.  

Constraints on operators and their nesting are specified via a Hyper-parameter \texttt{nested\_constraints} to the regressor. The following codes define a regressor targeting a function with only \texttt{log} and \texttt{sqrt}, allowing the single nesting form \texttt{log(sqrt(*))}.

\begin{widetext}
  \label{code:2}
  \begin{lstlisting}[language=Python, caption={Limit function nesting in PySR},captionpos=b]
  from pysr import PySRRegressor, TemplateExpressionSpec
  
  model = PySRRegressor(
      # Some other options 
      binary_operators=["*", "+", "-", "/"], # Allowing all binary operators
      unary_operators=["log", "sqrt"], # Forbid all unary functions except for log and sqrt
      nested_constraints={
          "sqrt": {"sqrt": 0, "log": 0},
          "log": {"sqrt": 1, "log": 0},
      },
      # Nesting constraints on operators. In this case, log(sqrt()) is the only permitted way for nesting. Nesting more than twice is prohibited either.
      # Some other options
  )
  \end{lstlisting} 
  \end{widetext}

For multi-variable case, this is not a standard usage of \textsc{PySR}. To achieve this, one should first declare a Hyper-parameter \texttt{combine} for \textsc{PySR} expression template:

\begin{widetext}
  \label{code:3}
  \begin{lstlisting}[language=Python, caption={Multi-variables template for indefinite integration in PySR},captionpos=b]
    from pysr import PySRRegressor, TemplateExpressionSpec
  
    template = TemplateExpressionSpec(
      expressions=["f"],
      variable_names=["x", "y", "dx", "dy"],
      combine="""
          fdx = D(f, 1)(x, y)
          fdy = D(f, 2)(x, y) 
          abs2(fdx - dx) + abs2(fdy - dy)
      """
      # In Multi-variable cases, one can manually define an error for each point. Here we apply Mean Square Error(MSE). 
    )
  \end{lstlisting}  
  \end{widetext}

This template allows \textsc{PySR} to compute derivatives for each variable and estimate regression errors pointwise. But this will fuse predicted values with targets and cause \textsc{PySR} cannot determine the error function for entire input series. Setting \texttt{elementwise\_loss} restricts the MSE calculation to predicted values only: 
  \begin{widetext}
  
  \begin{lstlisting}[language=Python,caption={Multi-variables regressor options in  PySR},captionpos=b]
    from pysr import PySRRegressor, TemplateExpressionSpec
  
    model = PySRRegressor(
      # Some other options 
      elementwise_loss="my_loss(predicted, target) = predicted",
      # Predicted value contains MSE of multi-variable derivatives. It is sufficient to present pointwise regression error.
      expression_spec = template,
      # Some other options
  )
  \end{lstlisting}  
  \end{widetext}
  
\end{document}